\documentclass[12pt]{iopart}

\usepackage{iopams} 
\usepackage{cite}
\usepackage{color}
\usepackage{graphicx}
\begin{document}

\title[Thin film superfluid optomechanics]{Thin film superfluid optomechanics}

\author{Christopher G. Baker, Glen I. Harris, David L. McAuslan, Yauhen Sachkou, Xin He and Warwick P. Bowen}

\address{Queensland Quantum Optics Laboratory, University of Queensland,
Brisbane, Queensland 4072, Australia}
\ead{c.baker3@uq.edu.au}
\vspace{10pt}
\begin{indented}
\item[]September 2nd, 2016
\end{indented}

\begin{abstract}
Excitations in superfluid helium represent attractive mechanical degrees of freedom for cavity optomechanics schemes. Here we numerically and analytically investigate the properties of optomechanical resonators formed by thin films of superfluid $^4$He covering micrometer-scale whispering gallery mode cavities. We predict that through proper optimization of the interaction between film and optical field, large optomechanical coupling rates $g_0>2\pi \times 100$ kHz and single photon cooperativities $C_0>10$ are achievable. Our analytical model reveals the unconventional behaviour of these thin films, such as thicker and heavier films exhibiting smaller effective mass and larger zero point motion. The optomechanical system outlined here provides access to unusual regimes such as $g_0>\Omega_M$ and opens the prospect of laser cooling a liquid into its quantum ground state.
\end{abstract}


\pacs{67.25.dt, 67.25.dp, 42.60.Da, 42.82.Et}
%
\vspace{2pc}
\noindent{\it Keywords}: cavity optomechanics, superfluidity, superfluid helium films, optical resonators, third sound, effective mass, liquids.
%
%
%

\section{Introduction}
\label{sectionintroduction}

The field of cavity optomechanics\cite{aspelmeyer_cavity_2014} focuses on the interaction between confined light and a mechanical degree of freedom. Optomechanical techniques enable an exquisite degree of control over the motion of micromechanical resonators, with successful examples including ground-state cooling \cite{chan_laser_2011} and squeezing of the mechanical motion of a resonator \cite{wollman_quantum_2015}. Recently, in a push to extend the realm of applications to biological systems, there has been 
a growing interest in the study of resonators immersed or interacting with liquids \cite{bahl2013brillouin, gil-santos_high-frequency_2015, fong2015nano}, or the use of liquids as resonators \cite{dahan_droplet_2016}.
In parallel, a special type of quantum liquid, namely superfluid helium, has also garnered significant attention for optomechanical applications, but for a different reason: the motivation here being the ultra-low optical and mechanical dissipation it can provide due to its reduced optical scattering \cite{seidel_rayleigh_2002} and absence of viscosity \cite{lorenzo_superfluid_2014, tilley1990superfluidity}; with both  properties being particularly desirable for quantum operations.

To date, the majority of superfluid optomechanics schemes have relied on bulk helium, with implementations taking for example the form of a gram-scale resonator coupled to a superconducting microwave resonator \cite{lorenzo_superfluid_2014}, a capacitively-detected superfluid Helmholtz resonator \cite{rojas_superfluid_2015} or a helium-filled fiber cavity \cite{kashkanova_superfluid_2016}.

In contrast, our group recently demonstrated an approach to superfluid optomechanics based on femtogram thin films of superfluid $^4$He condensed on the surface of a microtoroidal whispering gallery resonator \cite{harris_laser_2016, mcauslan_microphotonic_2016}. Leveraging the techniques of cavity optomechanics, we demonstrated real-time observation of the superfluid Brownian motion, laser cooling of the superfluid excitations \cite{harris_laser_2016}, as well as the  possibility to apply large optical forces at the microscale arising from the atomic recoil of superfluid helium flow \cite{mcauslan_microphotonic_2016}. 
However these devices, while exhibiting strong photothermal coupling, suffered from reduced radiation pressure coupling due to a poor overlap between the optical field and the superfluid mechanical excitations (known as third sound \cite{atkins_third_1959}, see sec. \ref{section2mechanics}).

In this work, we theoretically design a superfluid thin film resonator from the ground up, carefully optimizing the interaction between superfluid film and optical field (section \ref{section1optics}), in order to maximize the dispersive radiation-pressure optomechanical coupling.
We investigate three different resonator geometries (microdisk, annular microdisk and microsphere), and provide useful analytical expressions for the effective mass and zero-point motion of superfluid films (section \ref{section2mechanics}), as well as the scaling of optomechanical figures of merit with experimental parameters such as resonator dimensions and superfluid film thickness (section \ref{section3optomechanics}). Based upon this analysis,  we predict large optomechanical coupling rates $g_0>2\pi \times 100$ kHz and single photon cooperativities $C_0$ greater than 10 are achievable with experimentally accessible designs, as well as unconventional regimes such as $g_0$ greater than   the mechanical resonance frequency $\Omega_M$.

Superfluid thin films present a number of desirable properties: they are naturally self-assembling on the surface of any resonator due to a combination of ultra-low viscosity and attractive van der Waals forces \cite{tilley1990superfluidity}, they can be of extremely minute volume, with third sound detected in films only two monolayers thick \cite{scholtz_third_1974} and they offer a large degree of  tunability as their thickness and the frequency of the excitations they sustain can be swept in-situ over a large range simply by changing  the helium pressure in the sample chamber. In addition, third sound can exhibit strong Duffing non-linearities, be strongly coupled to quantized vortices \cite{yarmchuk_observation_1979, ellis_quantum_1993, gomez_shapes_2014, fonda_direct_2014} and interact with  electrons floating on the film \cite{yang_coupling_2016}. All together,
these properties underline the potential of superfluid thin films as a promising platform for cavity optomechanics.

%

\begin{figure}[h]
\centering
\includegraphics[width=\textwidth]{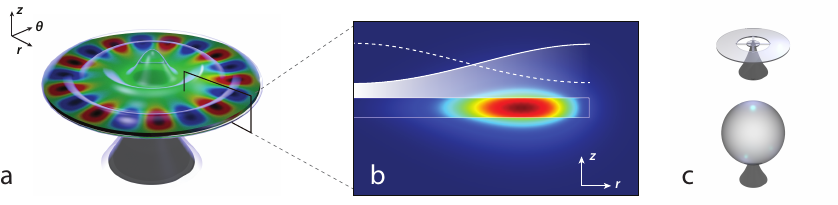}
\caption{(a) Artistic rendering of a disk-shaped optical resonator sustaining an optical WGM resonance (red and blue) covered in a superfluid helium thin film. (b) Radial cross-section showing a finite element method (FEM) simulation of the WGM field intensity. The oscillating superfluid wave on the top surface of the resonator (solid and dashed white lines) dispersively couples to the WGM via Eq. (\ref{Eqfreqshift}). (c) The case of an annular microdisk (top) and a microsphere resonator (bottom) are discussed separately in appendices A and B. }
\label{Figure1fig}
\end{figure}

\section{Optical field optimization}
\label{section1optics}

Figure \ref{Figure1fig} shows a schematic illustration of the optomechanical coupling scheme investigated in this paper. A circular whispering gallery mode (WGM) resonator \cite{matsko_optical_2006} is uniformly coated with a thin film of superfluid helium \cite{harris_laser_2016}. Acoustic waves in this thin film known as third sound \cite{atkins_third_1959, everitt_detection_1962, schechter_observation_1998} manifest as thickness variations which dispersively couple to the confined WGM via perturbations to its evanescent field. As shown in Fig. \ref{Figure1fig}(b), the fluctuating thickness of the superfluid in the vicinity of the WGM induced by a third sound wave modulates the amount of higher refractive index material in the WGM's near field, thereby changing the optical path length of the resonator. The frequency shift $\Delta \omega$ experienced by a WGM of resonance frequency $\omega_0$ due to the presence of the superfluid thin film is given by a perturbation theory approach \cite{anetsberger_near-field_2009}:

\begin{equation}
\frac{\Delta \omega}{\omega_0}=-\frac{1}{2}\, \frac{\int_{\mathrm{film}}\left(\varepsilon_{\mathrm{sf}}-1\right)\left\vert\vec{E}\left(\vec{r}\right)\right\vert^2 \mathrm{d}^3\vec{r}}{\int_{\mathrm{all}} \varepsilon_{r}\left( \vec{r}\right) \left\vert\vec{E}\left( \vec{r}\right)\right\vert^2 \mathrm{d}^3\vec{r}}
\label{Eqfreqshift}
\end{equation}
where $\vec{E}$ is the unperturbed WGM electric field calculated in the absence of superfluid, $\varepsilon_{r}\left( \vec{r}\right)$ is the relative permittivity and $\varepsilon_{\mathrm{sf}}=1.058$ is the relative permittivity of superfluid helium \cite{donnelly_observed_1998}. The  numerator integral is taken over the volume of the film while the denominator integral is taken over all space; Eq. (\ref{Eqfreqshift}) therefore essentially relates the electromagnetic (EM) energy `sensing' the perturbing element (the film) to the total EM energy in the mode. 
\begin{figure}
\centering
\includegraphics[width=\textwidth]{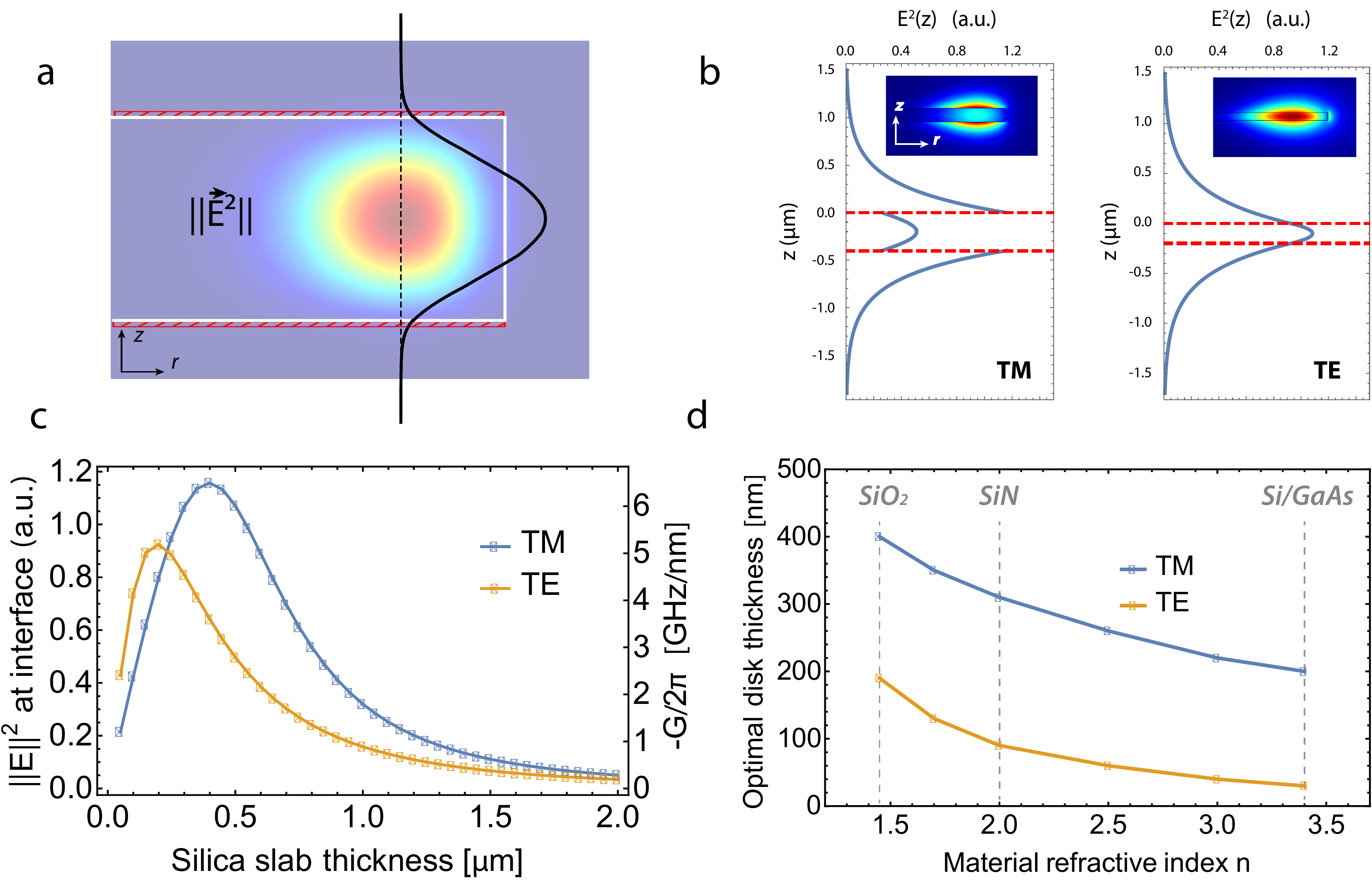}
\caption{(a) FEM simulation showing the radial cross-section of a TE WGM of wavelength 1.5 $\mu$m confined in a silica disk of 40 $\mu$m radius and 2 $\mu$m thickness. Overlayed in black is a plot of the $\left\vert\vec{E}\right\vert^2\left(z\right)$ mode profile along the dashed black line going through the center of the WGM. The red region indicates the superfluid film on top and bottom. (b) Vertical mode profiles $\left\vert\vec{E}\right\vert^2\left(z\right)$ for a 1.5 $\mu$m wavelength TM (resp. TE) WGM confined in a 400 nm (200 nm) thick, 40 $\mu$m radius disk. The dashed red lines mark the disk upper and lower boundaries. Inset: FEM radial cross-section of $|\vec{E}|^2$ for each WGM. (c) Value of $E^2$ at the interface calculated using the EIM, for TM (blue) and TE (orange) polarized modes. Right axis: optical frequency shift per nm of superfluid $G/2\pi$. (d) Indicative optimal disk thickness for maximal field at the interface and superfluid detection, as a function of disk material refractive index.}
\label{Figure2fig}
\end{figure}
Figure \ref{Figure2fig}(a) plots a FEM simulation of $|\vec{E}|^2$ for a transverse electric (TE) \cite{rosencher2002optoelectronics} WGM of wavelength $\lambda=1.5\,\,\mu$m confined in a typical silica disk of radius $R=40\,\mu$m  and 2 $\mu$m thickness \cite{kippenberg_fabrication_2003}. Overlayed in black is a plot of the vertical dependence of $\left\vert \vec{E}\right\vert^2$ along the dashed black line going through the center of the WGM. The field is mostly confined within the silica resonator, and has low intensity on the top and bottom interface where it is in contact with the superfluid film (red dashes), resulting in weak detection sensitivity/optomechanical coupling through Eq. (\ref{Eqfreqshift}). This limitation can be overcome by a proper choice of disk thickness and WGM polarization so as to maximize the electric field at the interface, as discussed in the following. Figure \ref{Figure2fig}(b) shows the same vertical mode profile for a transverse magnetic (TM) polarized WGM in a 400 nm thick silica disk of identical radius, and for a TE WGM  confined in a 200 nm thick disk. Reducing the thickness of the disk pushes the field out of the resonator, increasing $\left\vert \vec{E}\right\vert^2$ over the superfluid region, such that any fluctuation in the film thickness results in a much larger frequency shift of the WGM. Figure \ref{Figure2fig}(c) provides a more systematic investigation of this mechanism. 
We employ the effective index method (EIM) \cite{rosencher2002optoelectronics} to calculate the mode profile for a TE and TM guided wave inside a slab waveguide of varying thickness (normalized such that $\int_{-\infty}^{\infty}\varepsilon_r E^2 \mathrm{d}z=1$), recording for each thickness the value of $E^2$ at the interface. For each polarization there is an optimal thickness which maximizes $E^2$ at the interface: too thick and the field is mostly confined within the resonator, too thin and the field becomes very delocalized along z and its value at the interface drops again.
From this analysis, we obtain the optimal silica disk thickness for TM (TE) WGMs as approximately 400 nm (200 nm).
Next, we repeat the same analysis for different values of the refractive index of the slab, while tracking the optimal thickness for TE and TM WGMs. These results are summarized in Fig. \ref{Figure2fig}(d) and show for instance that for a high refractive index material such as silicon or gallium arsenide, a disk thickness of $\sim 200$ nm is optimal for TM polarized WGMs.
Since these results are based on the EIM, they start to lose accuracy for strongly confining geometries (R on the order of a few $\lambda$); nevertheless they provide a useful starting point for designing optimized structures. The TM polarized WGMs, with dominant field component $E_z$ normal to the upper and lower interfaces, provide a step increase in the field outside the disk (due to the continuity of $\varepsilon_r E_z$), as shown in Fig. \ref{Figure2fig}(b), and are therefore more sensitive to the superfluid \cite{borselli_beyond_2005}.

Since the evanescent field decays along $z$ with characteristic length on the order of hundreds of nanometers (Fig. \ref{Figure2fig}(b)), the change in $\left\vert \vec{E}\right\vert^2$ over the typical $\Delta z=1$ to 30 nm thickness of the film can safely be neglected. Equation (\ref{Eqfreqshift}) can therefore be rearranged to give the optomechanical optical resonance frequency shift per unit displacement $G=\Delta \omega_0/\Delta z$ \cite{aspelmeyer_cavity_2014, bowen2015quantum,baker_photoelastic_2014}:
\begin{equation}
G=\frac{\Delta \omega_0}{\Delta z}=-\frac{\omega_0}{2}\, \frac{\int_{\mathrm{interface}}\left(\varepsilon_{\mathrm{sf}}-1\right)\left|\vec{E}\left(\vec{r}\right)\right|^2 \mathrm{d}^2\vec{r}}{\int_{\mathrm{all}} \varepsilon_{r}\left( \vec{r}\right) \left|\vec{E}\left( \vec{r}\right)\right|^2 \mathrm{d}^3\vec{r}}
\label{EqG}
\end{equation}
where the numerator integral is now a surface integral over the resonator top interface\footnote{For simplicity we will in the following consider excitations on the top surface of the resonator, although the treatment follows the same approach for the bottom surface, but with different boundary conditions for the third sound wave due to the presence of the pedestal.}. We employ the one dimensional version of Eq. (\ref{EqG}) ($G=-\frac{\omega_0}{2}\,$ $ \left(\varepsilon_{\mathrm{sf}}-1\right)E^2\left(0\right)/\int \varepsilon_{r}\left( z\right) E^2\left( z\right) \mathrm{d}z$) 
to provide $G/2\pi$ as a function of resonator thickness in the right axis of Fig. \ref{Figure2fig}(c). As for double-disk optomechanical resonators \cite{jiang_high-q_2009}, G is independent of resonator radius and does also not depend on superfluid thickness. Proper choice of resonator thickness is important, resulting for instance in a 20-fold improvement in G for TM modes when going from a 2 $\mu$m to a 0.4 $\mu$m thick silica disk. Through this optimization, it is possible to reach large values of G upwards of 6 GHz/nm, despite the challenge posed by superfluid helium's optical properties being very close to those of vacuum ($\varepsilon_r=1.058$, $n=1.029$) \cite{donnelly_observed_1998}. This is achieved in part thanks to the perfect spatial overlap provided by the self-assembling nature of the superfluid film: these predicted coupling rates are for instance nearly three orders of magnitude larger than those obtained in experiments in which a silicon nitride string was approached in a microtoroid's near field  \cite{anetsberger_near-field_2009}.
Note additionaly that unlike the previously mentioned scheme in which a perturbing element of constant volume is approached in the near field of a resonator \cite{anetsberger_near-field_2009,favero_fluctuating_2009,cole2015evanescent,schilling_near-field_2016}, this detection approach does not rely on an electric field gradient along the $z$ direction, as we are instead detecting a perturbing element of changing volume.

It is interesting to examine whether large coupling rates can also be achieved for superfluid modes confined to the vertical sidewall of the WGM. For a circular WGM cavity, the sensitivity to a change in radius is given by $G_{\mathrm{radial}}\simeq\frac{\omega_0}{R}$ \cite{ding2010high}. From Eq. (\ref{EqG}), but integrated along the vertical, rather than top boundary, it can be shown that fluctuations in the thickness of the superfluid film on the vertical boundary would translate to $G_{\mathrm{vertical}}/2\pi\simeq\frac{\omega_0}{R}\left(\frac{1-\varepsilon_\mathrm{sf}}{1-\varepsilon_\mathrm{SiO_2}}\right)/2\pi$
 $\simeq$ 0.25 GHz/nm with the above parameters.

Here we have discussed the sensitivity to thickness fluctuations affecting  either the top, bottom or vertical boundaries of the resonator. Naturally, a variation in the mean thickness of the film would produce a frequency shift of $2 G+ G_{\mathrm{vertical}} > 10$ GHz/nm. This means a change in film thickness of 10 pm (i.e. $\frac{1}{36}$\textit{th} of a helium monolayer \cite{sabisky_verification_1973}) would be sufficient to shift a WGM resonance with $Q=2\times 10^6$ by one linewidth, thereby providing an ultra-precise independent means to optically characterize the superfluid film thickness, a significant improvement over capacitive detection schemes commonly used in the superfluid community \cite{keller_thickness_1970, schechter_observation_1998}.

\section{Superfluid third sound modes}
\label{section2mechanics}
Third sound waves \cite{atkins_third_1959, everitt_detection_1962, scholtz_third_1974,ellis_observation_1989, schechter_observation_1998, harris_laser_2016} are a type of  excitation unique to superfluid thin films which manifest as thickness fluctuations with a restoring force provided by the van der Waals interaction; they are somewhat analogous to water waves (where the restoring force is gravity) \cite{phillips1969dynamics}, see Fig. \ref{Figure3fig}(a). 
Third sound propagates at a speed $c_3$ given by: \cite{tilley1990superfluidity}
\begin{equation}
c_3=\sqrt{3\frac{\rho_s}{\rho}\frac{\alpha_{\mathrm{vdw}}}{d^3}},
\label{Eqc3}
\end{equation}
with $\rho_s/\rho$ the ratio of superfluid to total fluid density \cite{tilley1990superfluidity}, $\alpha_{\mathrm{vdw}}$ the van der Waals coefficient characterizing the strength of the attractive force between the helium atoms and the substrate, and $d$ the superfluid film mean thickness. The van der Waals coefficients for various resonator materials are provided in Table \ref{Tablevanderwaals}. The $1/d^3$ dependency in $c_3$ neglects the retardation effects in the van der Waals potential and is a reasonable first order approximation for films 0 to 30 nm thick \cite{sabisky_verification_1973, enss_low-temperature_2005} which we will consider here.  
\begin{table}
\centering
\begin{tabular}{l c c}
\hline
Material & $\alpha_{\mathrm{vdw}}$  & source \\
\hline
Silica & $2.6 \times 10^{-24}$ m$^5$s$^{-2}$ & \cite{sabisky_onset_1973,scholtz_third_1974}\\
CaF$_2$ & $2.2 \times 10^{-24}$ m$^5$s$^{-2}$ & \cite{sabisky_onset_1973}\\ 
Silicon & $3.5 \times 10^{-24}$ m$^5$s$^{-2}$ & \cite{sabisky_onset_1973}\\
MgO & $2.8 \times 10^{-24}$ m$^5$s$^{-2}$ & \cite{sabisky_verification_1973}\\
\hline
\end{tabular}
\caption{Van der Waals coefficients for a few resonator materials.}
\label{Tablevanderwaals}
\end{table}
The disk geometry, in addition to providing optical confinement to the WGMs, also confines third sound excitations localized on the top surface, giving rise to third sound resonances.  The shape of these resonant modes is dictated by the confining geometry and, for circular resonators, these take the form of Bessel modes as shown in Fig. \ref{Figure3fig}(a) \cite{schechter_observation_1998,ellis_observation_1989, ellis_quantum_1993}. The third sound mode profile $\eta_{m,n}$ describes the out-of-plane deformation of the superfluid surface for the ($m$; $n$) mode
as a function of time $t$ and polar coordinates $r$ and $\theta$:
\begin{equation}
\eta_{m,n}\left(r,\theta, t\right)= A_{m,n}\,J_m\left(\zeta_{m,n} \frac{r}{R}\right)\cos\left(m \theta\right)\sin\left(\Omega_M t\right),
\label{EqBessel}
\end{equation}
where $m$ and $n$ are respectively the azimuthal and radial mode numbers, A the mode amplitude, $J_m$ the Bessel function of the first kind of order $m$, $\Omega_M=(\zeta_{m,n} c_3)/R$ the mode frequency and $\zeta_{m,n}$ a frequency parameter depending on the mode order and the boundary conditions, see Table \ref{TablezeroesBessel}. In the following we only focus on the rotationally invariant modes ($m$=0), as these are the ones with largest optomechanical coupling \cite{ding2010high}.
\begin{table}
\centering
\begin{tabular}{l c c c c c c c c c c }
\hline
 & n=1 & n=2 & n=3 & n=4 & n=5 & n=6 & n=7 & n=8 & n=9 & n=10 \\
\hline
Free & 3.832 &7.016& 10.174 &13.324 &16.471 & 19.616 & 22.760 & 25.904 & 29.047 & 32.19 \\
Fixed  &2.405 &  5.520 &8.654 &11.792 &14.931 &18.071 & 21.212& 24.353 & 27.49 & 30.63\\
\hline
\end{tabular}
\caption{First eight values of the frequency parameter $\zeta_{0,n}$ for fixed and free boundary conditions. In the case of fixed (free) boundary conditions, these correspond to the zeroes of $J_0$ ($J_0'$).}
\label{TablezeroesBessel}
\end{table}
\subsection{Boundary conditions}
\begin{figure}
\centering
\includegraphics[width=\textwidth]{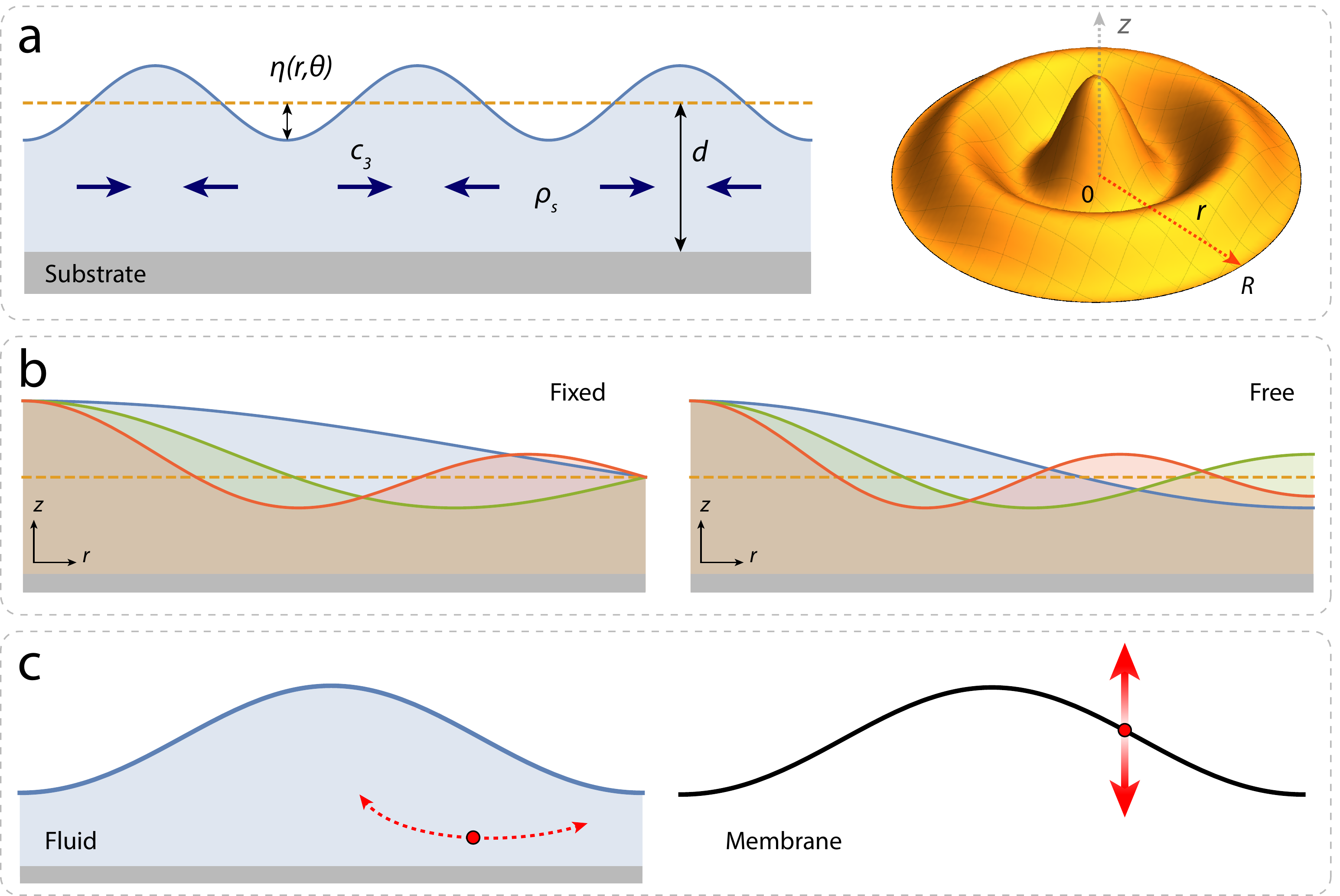}
\caption{\textbf{Superfluid third sound}. (a) Left: schematic illustration of a superfluid third sound wave with profile $\eta\left(\vec{r}\right)$ on a film of mean thickness d (dashed orange line). The normal fluid component \cite{tisza_theory_1947} is viscously clamped to the surface, while the superfluid component $\rho_s$ oscillates mostly parallel to the substrate (blue arrows). Right: Plot of the surface profile for the ($m$=0; $n$=3) Bessel mode with free boundary conditions. (b) Radial  profile $\eta\left(r\right)$ along the dashed red line in (a) for the ($m$=0; $n$=1) --blue--, ($m$=0; $n$=2) --green-- and ($m$=0; $n$=3) --red-- Bessel modes with fixed (left) and free (right) boundary conditions. (c) Comparison between the trajectory described by a `particle' in a fluid (left) and a solid membrane (right).}
\label{Figure3fig}
\end{figure}

The mode profiles for the first three rotationally invariant third sound modes with fixed ($\eta(R)=0$) and free ($\partial_r\eta(R)=0$) boundary conditions \cite{schechter1999third} are shown in Fig. \ref{Figure3fig}(b). The free boundary condition is also known as the `no flow' boundary condition, as it requires the radial velocity of the superfluid flow to be 0 at $r=R$ \cite{schechter1999third}, and is therefore volume conserving. On the contrary, the fixed boundary condition does not conserve volume (particularly visible for the ($m$=0; $n$=1) mode), and therefore requires significant flow across the confining boundary, in the incompressible limit. In circular $^3$He third sound resonators a crossover from free to fixed boundary conditions has been observed for films thicker than $\sim 200$ nm \cite{vorontsov_spectrum_2004,schechter1999third}. For our resonator design, with thin films and near atomically sharp `knife-edge like' \cite{shirron_suppression_1998}  microfabricated boundaries, we expect minimal third sound driven flow to occur between the disk's upper and lower surfaces. This implies free boundary conditions for the third sound wave, consistent with those observed in circular $^4$He third sound resonators most similar to our design
\cite{ellis_observation_1989,ellis_quantum_1993}.
\subsection{Third sound mode effective mass}
\label{subsectioneffectivemass}

Calculating the optomechanical single photon coupling strength $g_0$ and  single photon cooperativity $C_0$ -- useful figures of merit of the optomechanical system \cite{aspelmeyer_cavity_2014} (see Eq. (\ref{Eqg0-C0})) -- requires the third sound mode effective mass $m_{\mathrm{eff}}$.
\begin{equation}
g_0=G\, x_{\mathrm{zpf}}=G \sqrt{\frac{\hbar}{2\, m_{\mathrm{eff}} \,\Omega_M}} \qquad\qquad C_0=\frac{4\, g_0^2}{\kappa\, \Gamma_M}\,.
\label{Eqg0-C0}
\end{equation}

In continuum mechanics, the effective mass at reduction point $\vec{A}$ is obtained by reducing the system to a point mass $m_{\mathrm{eff}}$ moving with velocity $v(\vec{A})$ possessing the same kinetic energy $E_k$ as the original system, that is $m_{\mathrm{eff}}=\frac{2 E_k}{v^2_A}$, or:

\begin{equation}
m_{\mathrm{eff}}=\frac{\int_V \rho\, v^2\left(\vec{r}\right)\mathrm{d}^3\left(\vec{r}\right)}{v^2\left(\vec{A}\right)}\,.
\label{Eqmeffgeneral}
\end{equation}
For rotationally invariant modes of a thin solid circular resonator of thickness $d$, this leads to the well known expression for the effective mass of a point on the resonator boundary \cite{wang_1.156-ghz_2004}:
\begin{equation}
m_{\mathrm{eff\,\, solid}}=2\pi\rho d\,\frac{ \int_0^R r\, \eta^2\left(r\right)\mathrm{d}r}{\eta^2\left(R\right)}\,.
\label{Eqmeffsolid}
\end{equation}
Going from Eq. (\ref{Eqmeffgeneral}) to Eq. (\ref{Eqmeffsolid}) assumes the velocity $v\left(\vec{r}\right)$ (and density) do not depend on the $z$ coordinate, which is valid for both in- and out-of-plane mechanical modes. 
Here, in order to circumvent the question of the distribution of superfluid velocity \cite{atkins_third_1959} and density \cite{scholtz_third_1974} below the film surface, we use the equipartition theorem to replace the kinetic energy term in Eq. (\ref{Eqmeffgeneral}) with the van der Waals potential energy stored in the deformation of the film surface. In analogy to gravity waves \cite{phillips1969dynamics}, this energy is given by:
\begin{equation}
\fl \eqalign{
E_{\mathrm{pot}}&=\rho\int_{\theta=0}^{2\pi}\int_{r=0}^R\left(\int_{z=0}^{d+\eta\left(r, \theta\right)} U\left(z\right)\mathrm{d}z\right)r \,\mathrm{d}r \,\mathrm{d}\theta -\rho\int_{\theta=0}^{2\pi}\int_{r=0}^R\left(\int_{z=0}^d U\left(z\right)\mathrm{d}z\right)r \,\mathrm{d}r \,\mathrm{d}\theta \\
&=\rho\int_{0}^{2\pi}\int_{0}^R\left(\int_d^{d+\eta\left(r, \theta\right)} U\left(z\right)\mathrm{d}z\right)r \,\mathrm{d}r \,\mathrm{d}\theta,}
\label{EqEpot1}
\end{equation}
where $U\left(z\right)$ is the energy per unit mass of the film due to the van der Waals potential \cite{tilley1990superfluidity}:
\begin{equation}
U\left(z\right)=-\frac{\alpha_{\mathrm{vdw}}}{z^3}\,.
\label{EqUvdw}
\end{equation}
In the limit of small amplitude surface oscillation $\eta \ll d$, we obtain: 
\begin{equation}
\int_d^{d+\eta\left(r, \theta\right)} U\left(z\right)\mathrm{d}z= -\frac{\alpha_{\mathrm{vdw}}\,\eta\left(r, \theta\right) }{d^3}+ \frac{3\,\alpha_{\mathrm{vdw}} \,\eta^2\left(r, \theta\right)}{2\, d^4}\,.
\end{equation}
Therefore Eq. (\ref{EqEpot1}) becomes for rotationally invariant modes:
\begin{equation}
E_{\mathrm{pot}}=2\pi\rho\int_0^R \rmd r\, r \,\left(- \frac{\alpha_{\mathrm{vdw}}\,\eta\left(r\right) }{d^3}+\frac{3\, \alpha_{\mathrm{vdw}} \,\eta^2\left(r\right)}{2\, d^4} \right).
\end{equation}
As mentioned previously, free (`no flow') boundary conditions are volume conserving ($\int_0^R r \,\rmd r\, \eta\left(r\right)=0$), therefore we obtain for the effective mass of a point on the film surface at $r$=$R$:
\begin{equation}
m_{\mathrm{eff}}=\frac{2 E_{\mathrm{pot}}}{v^2\left(R\right)}=\frac{6\pi \, \rho \,  \alpha_{\mathrm{vdw}} \, d^{-4}\int_0^R r \, \eta^2\left(r\right)\,\rmd r}{\eta^2\left(R\right)\,\Omega_M^2}\,.
\end{equation}
Finally, using $\Omega_M=\frac{\zeta c_3}{R}$ and Eq. (\ref{Eqc3}) we find:
\begin{equation}
m_{\mathrm{eff}}=\left(\frac{\rho}{\rho_s}\right)\left(\frac{R}{d}\right)^2 \frac{1}{ \zeta^2}\,\times\,2\pi\rho \,d \,\frac{ \int_0^R r \,\eta^2\left(r\right)\,\mathrm{d}r}{\eta^2\left(R\right)} \,.
\label{Eqmeffsuperfluid}
\end{equation}
Here we recognize the effective mass of the solid case (Eq. (\ref{Eqmeffsolid})), multiplied by a prefactor  proportional to $(R/d)^2$. This means that while for a solid such as a circular membrane $m_{\mathrm{eff}}$ scales as expected as $R^2 d$ (like the real mass), for a third sound  wave on a superfluid film $m_{\mathrm{eff}}$ scales as $R^4/d$  ---with thicker and heavier films therefore making for lighter resonators possessing larger zero point motion. This radically different scaling can be understood by considering the microscopic motion of a `particle' of the resonator in both cases, as illustrated in Fig. \ref{Figure3fig}(c). Indeed, while the surface deformation in each case is governed by the same mathematical equation (Eq. (\ref{EqBessel})), a particle in the solid describes an essentially vertical motion while that in a fluid an extremely flattened near horizontal trajectory. As the useful displacement for optomechanical coupling is in the $z$ direction, the horizontal particle excursion ($\propto R$) and horizontal kinetic energy are `wasted' and appear as a penalty term in the effective mass.

The $R^4$ dependence of $m_{\mathrm{eff}}$ underscores the dramatic gains achieved by going towards smaller microfabricated third sound resonators. Indeed, going from a centimeter-scale third sound resonator \cite{ellis_quantum_1993} to a 40 micron radius resonator such as outlined here and demonstrated in Ref. \cite{harris_laser_2016} affords  ---with otherwise identical parameters--- a $5\times10^8$ reduction in effective mass and identical boost in cooperativity (Eq. (\ref{Eqg0-C0})).

Note interestingly that we recover exactly the same $R^4/d$ effective mass scaling if we consider a gravitational wave in a normal liquid (by substituting the gravitational potential $g\,z$ in  Eq. (\ref{EqUvdw})), and consider the shallow water limit ($\lambda \gg d$) where the speed of sound $c=\sqrt{g\, d}$ only depends on liquid height \cite{phillips1969dynamics}.

\section{Optomechanical coupling}
\label{section3optomechanics}

In this section we evaluate the performance of superfluid thin films as optomechanical resonators and successively address how this performance is influenced by resonator dimensions, film thickness and mechanical mode order.

\subsection{Influence of resonator radius}

\begin{figure}
\centering
\includegraphics[width=\textwidth]{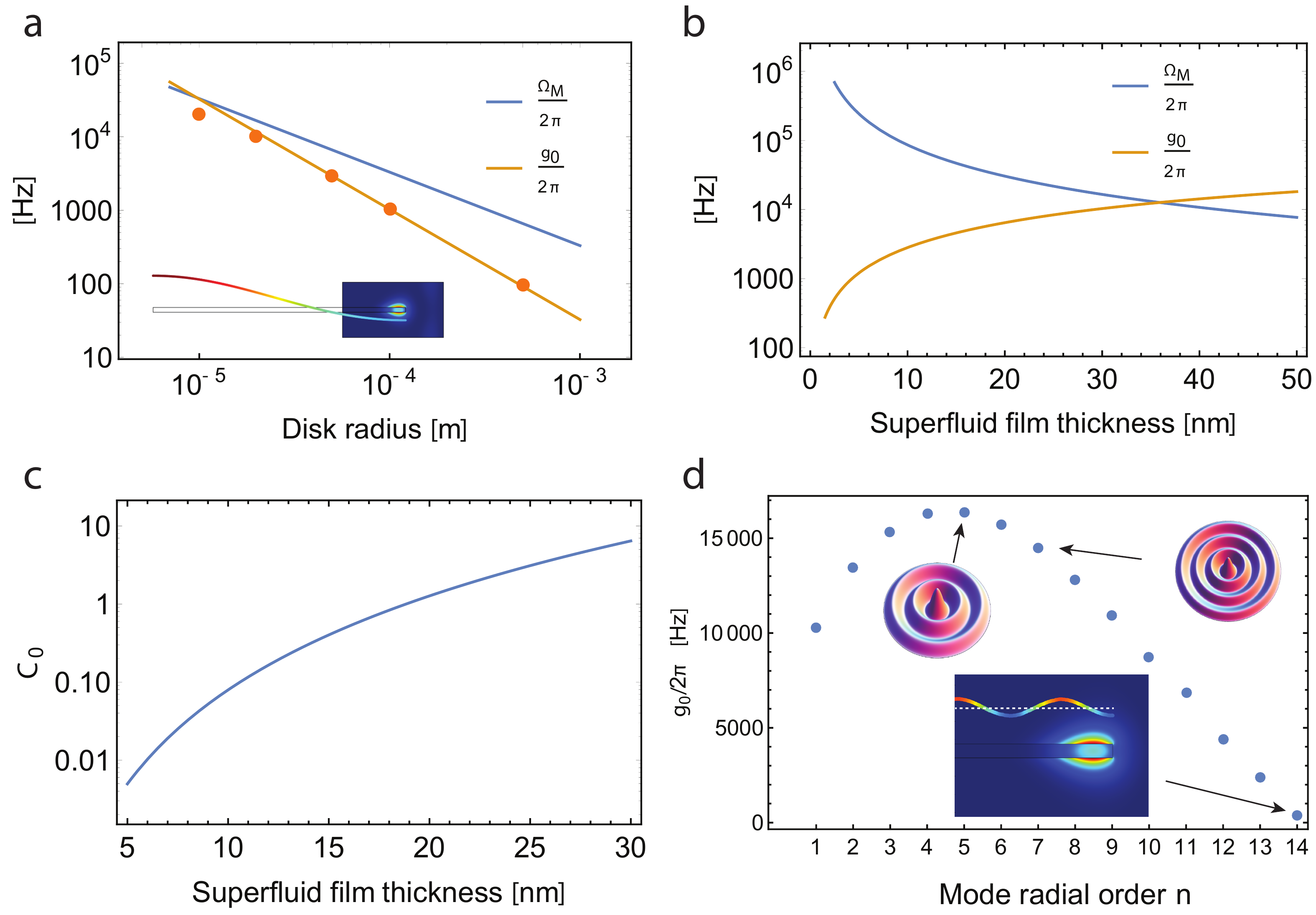}
\caption{\textbf{Optomechanical parameters}.(a) Single photon optomechanical coupling strength $g_0/2\pi$ and mechanical frequency $\Omega_M/2\pi$ for the fundamental ($m$=0; $n$=1) third sound mode as a function of resonator radius, for a $d=30$ nm thick superfluid film. Solid line: analytical formula, points: individual FEM simulations. Inset: FEM simulation displaying the WGM overlayed with the third sound mode displacement profile (colored line). (b) $\Omega_M/2\pi$ (blue) and  $g_0/2\pi$ (orange) as a function of film thickness for a $R=20$ $\mu$m disk. (c) Predicted single photon optomechanical cooperativity $C_0$ as a function of film thickness $d$, for a $R=20$ $\mu$m disk. (d) Influence of third sound radial mode order $n$ on $g_0/2\pi$, for $R=20$ $\mu$m and $d=30$ nm. Inset: mechanical surface deformation profiles and FEM simulation showing the WGM mode overlayed with the displacement profile of the ($m$=0; $n$=14) third sound mode. All results are for free boundary conditions.}
\label{Figure4fig}
\end{figure}

Figure \ref{Figure4fig}(a) plots the dependence of third sound frequency $\Omega_M/2\pi$ and $g_0/2\pi$ on resonator radius for the fundamental ($m$=0; $n$=1) third sound mode on a 30 nm thick superfluid film. The solid orange line corresponds to the value of $g_0$ given by Eq. (\ref{Eqg0-C0}), employing the previously determined value of $m_{\mathrm{eff}}$ (Eq. (\ref{Eqmeffsuperfluid})) and assuming a constant $G/2\pi=6.6$ GHz/nm (see section \ref{section1optics}). This is a good assumption for disk radii above $\sim20
\,\mu$m, as the micron sized radial extension of the WGM is small compared to R, and the superfluid displacement is therefore essentially constant over the optical mode (see inset). For smaller radii (and higher order mechanical modes ---see section \ref{subsectionradialmodeorderinfluence}) the mode overlap between the optical field and the third sound displacement field needs to be taken into account when calculating $g_0$:
\begin{equation}
g_0=-\frac{\omega_0}{2}\, \frac{\int_{\mathrm{interface}}\,q  \left(\varepsilon_{\mathrm{sf}}-1\right)\left|\vec{E}\left(\vec{r}\right)\right|^2 \mathrm{d}^2\vec{r}}{\int_{\mathrm{all}} \varepsilon_{r}\left( \vec{r}\right) \left|\vec{E}\left( \vec{r}\right)\right|^2 \mathrm{d}^3\vec{r}}\,.
\label{Eqg0}
\end{equation}
Here $q=\frac{\eta\left(r\right)}{\eta\left(R\right)}\,x_{\mathrm{zpf}}$ is the superfluid displacement profile normalized to $x_{\mathrm{zpf}}$ at $R$. The orange dots correspond to the results of individual FEM simulations using Eq. (\ref{Eqg0}) with TM WGMs on a 380 nm thick silica disk. The analytical expression is in good agreement with the FEM simulation down to $R\sim 20$ $\mu$m ($<10$ \% error), below which it overestimates $g_0$. Since the mechanical frequency scales as $R^{-1}$ and the zero point motion as $R^{-3/2}$, $g_0$ increases faster than $\Omega_M$ as R is reduced, and for $R\leq20$ the system enters the uncommon optomechanical regime of $g_0>\Omega_M$. Table \ref{Tablescalingparameters} summarizes the relevant scaling parameters for a disk resonator based on Eqs. (\ref{Eqc3}), (\ref{EqBessel}), (\ref{Eqg0-C0}) and (\ref{Eqmeffsuperfluid}).

\begin{table}
\centering
\begin{tabular}{l c c c c }
& & $R$ & $d$ & $\zeta$  \\
\hline
$m$ &$\propto$& $R^2$ & $d$ & $-$\\
$m_{\mathrm{eff}}$ &$\propto$& $R^4$ & $d^{-1}$ & $\zeta^{-2}$ \\
$\Omega_M$ &$\propto$& $R^{-1}$ & $d^{-3/2}$ & $\zeta$ \\
$x_{\mathrm{ZPF}}$  &$\propto$&$R^{-3/2}$ & $d^{5/4}$& $\zeta^{1/2}$\\
$g_0$ & $\propto$&$\dagger$ & $d^{5/4}$& $\dagger$\\
\hline
\end{tabular}
\caption{Scaling of experimental parameters with resonator radius $R$, film thickness $d$ and mode order $\zeta$. Table should be read horizontally. Symbols $-$ and $\dagger$ respectively denote no dependence and a non-monotonous dependence. }
\label{Tablescalingparameters}
\end{table}
\subsection{Influence of superfluid film thickness}
\label{sectionInfluenceoffilmthickness}

Figure \ref{Figure4fig}(b) plots the dependence of $\Omega_M/2\pi$ and $g_0/2\pi$ on superfluid film thickness, for a $R=20$ um resonator. Since $\Omega_M$ scales as $d^{-3/2}$ and $g_0$ as $d^{5/4}$ (see Table \ref{Tablescalingparameters}), both parameters start with extremely dissimilar values for thin films and evolve towards one another as $d$ increases. In this particular case, $g_0\simeq\Omega_M \simeq 2\pi\times 13.5$ kHz for $d=34$ nm. 
We take into account the WGM $\left\vert\vec{E}\right\vert^2$ field decay as the film gets thicker, but this is only a minor correction for the thin films we consider here.
Next we plot in Fig. \ref{Figure4fig}(c) the dependence of the single photon cooperativity $C_0$ (Eq. (\ref{Eqg0-C0})) on $d$. For this estimation we consider an optical loss rate $\kappa/2\pi=20$ MHz corresponding to an optical Q of $10^7$ as demonstrated in thin silica disks of identical radius\footnote{The presence of the superfluid helium film around the resonator does not adversely affect the optical Q due to superfluid helium's ultralow optical absorption in the infrared \cite{harris_laser_2016, kashkanova_superfluid_2016}.} \cite{schilling_near-field_2016}, and a conservative estimate for the mechanical $Q_M=\Omega_M/\Gamma_M$ of 4000, as demonstrated in our previous work  with microtoroid resonators \cite{harris_laser_2016}. (Note that third sound dissipation rates several orders of magnitude below these values have already been demonstrated \cite{ellis_observation_1989}). The predicted cooperativity displays a strong dependence on film thickness, reaching large values above unity, with $C_0=6$ for $d=30$ nm.
This value -- on par with the state-of-the-art in optomechanical systems \cite{chan_laser_2011, yuan_large_2015} -- would represent a significant increase in performance compared to existing superfluid optomechanics systems; it corresponds for instance to an over four orders of magnitude increase over recently demonstrated superfluid helium filled fiber cavities \cite{kashkanova_superfluid_2016}.

\subsection{Influence of third sound mode radial order $n$}
 \label{subsectionradialmodeorderinfluence}

The effective mass of a third sound mode is inversely proportional to $\zeta^2$, as shown in equation (\ref{Eqmeffsuperfluid}). This relationship arises because as $\zeta$ increases, the ratio of vertical to horizontal superfluid motion becomes more favorable, as the distance  between the peaks and the troughs of the third sound wave is reduced (see figure \ref{Figure3fig}). Higher order radial third sound modes (with higher $\zeta$; see table \ref{TablezeroesBessel}) therefore exibit  lower $m_{\mathrm{eff}}$ and larger $x_{\mathrm{zpf}}$ (see Table \ref{Tablescalingparameters}). Note that this is the opposite behaviour to that for a solid membrane, in which $x_{\mathrm{zpf}}$ decreases with increasing mode order. Figure \ref{Figure4fig}(d) plots results of FEM simulations of $g_0$ versus third sound mode radial order $n$, obtained through equation (\ref{Eqg0}) for a $R=20$ $\mu$m disk with $d=30$ nm. It reveals two competing trends. First, and initially dominating, is the increase in $g_0$ due to the increase in $x_{\mathrm{zpf}}$. Second, for higher $n$, the third sound displacement becomes oscillatory over the WGM (see inset)
leading to a dramatically reduced overlap integral, see equation (\ref{Eqg0}). In this particular case $g_0$ reaches its maximal value of $2\pi\times 16$ kHz for $n=5$ and $\Omega_M/2\pi=71$ kHz. The optimal radial order will naturally depend on device dimensions, with larger $R$ leading to a larger optimal $n$. These higher order modes provide a twofold benefit: beyond the higher $g_0$, they have a higher frequency and therefore  exhibit a lower thermal phonon occupancy $\bar{n}$ for a given bath temperature. Starting from a base temperature of 20 mK, on the order of $n_c=2\times 10^3$ intracavity photons would be sufficient to feedback cool \cite{lee_cooling_2010,mcauslan_microphotonic_2016} this mode into its quantum ground state \cite{chan_laser_2011}.

\subsection{Miniaturized third sound resonators}
\label{sectionminiaturizedresonators}

Finally we briefly address the potential of micrometer-sized WGM resonators made of high refractive index semiconductors such as silicon \cite{sun_high-q_2012,borselli_beyond_2005} or gallium arsenide \cite{ding_wavelength-sized_2011,gil-santos_high-frequency_2015} for thin-film superfluid optomechanics. The optimal resonator thicknesses given in figure \ref{Figure2fig}(d) are chosen to maximize the WGM deconfinement, resulting in low WGM effective indices, and therefore do not lend themselves to wavelength-sized  radii without incurring significant bending losses \cite{parrain_origin_2015}. For this reason we consider thicker more confining disks for this application, such as 200 nm thickness for TE modes \cite{ding_wavelength-sized_2011}. This trade-off between sensitivity and optical Q results in a lower $G/2\pi$ on the order of 1 GHz/nm. The $\Omega_M/2\pi=330$ kHz (m=0; n=1) third sound mode of a 30 nm thick film confined on top of a R=1 $\mu$m disk has a zero point motion $x_{\mathrm{zpf}}=1.6\times 10^{-13}$ m at the periphery and a large optomechanical coupling rate reaching up to $g_0=2\pi \times 136$ kHz. Because of the lower optical Q of these resonators however, the expected cooperativity $C_0$ is on par with that expected for larger silica resonators.

\section{Conclusion}
\label{sectionconclusion}

We have investigated the potential of thin films of superfluid $^4$He covering micrometer-scale whispering gallery mode cavities as optomechanical resonators. Our analysis predicts large optomechanical coupling and cooperativities are achievable, and provides useful tools for the design of third-sound optomechanical resonators. Furthermore, beyond their sole merits for `conventional' optomechanics, the ability to engineer interactions between third sound phonons and quantized vortices \cite{yarmchuk_observation_1979, ellis_quantum_1993, gomez_shapes_2014, fonda_direct_2014} as well as electrons \cite{yang_coupling_2016},
combined with the ability to control superfluid flow on-chip  \cite{mcauslan_microphotonic_2016} and generate long-lived persistent flows \cite{ekholm_studies_1980}, makes this a promising platform with varied applications such as ground-state cooling of a liquid, on-chip inertial sensing \cite{sato_superfluid_2012}, single-photon optomechanics \cite{nunnenkamp_single-photon_2011, tang_prospect_2014} and the study of the dynamics of strongly interacting quantum fluids.


%

\section*{Acknowledgments}
This work was funded by the Australian Research Council through the Centre of Excellence for Engineered Quantum Systems (EQuS, CE110001013); W.P.B. acknowledges the Australian Research Council Future Fellowship FT140100650.

\clearpage

\section{Appendix A: Annular microdisk}
\label{sectionannulardisk}

\begin{figure}
\centering
\includegraphics[width=\textwidth]{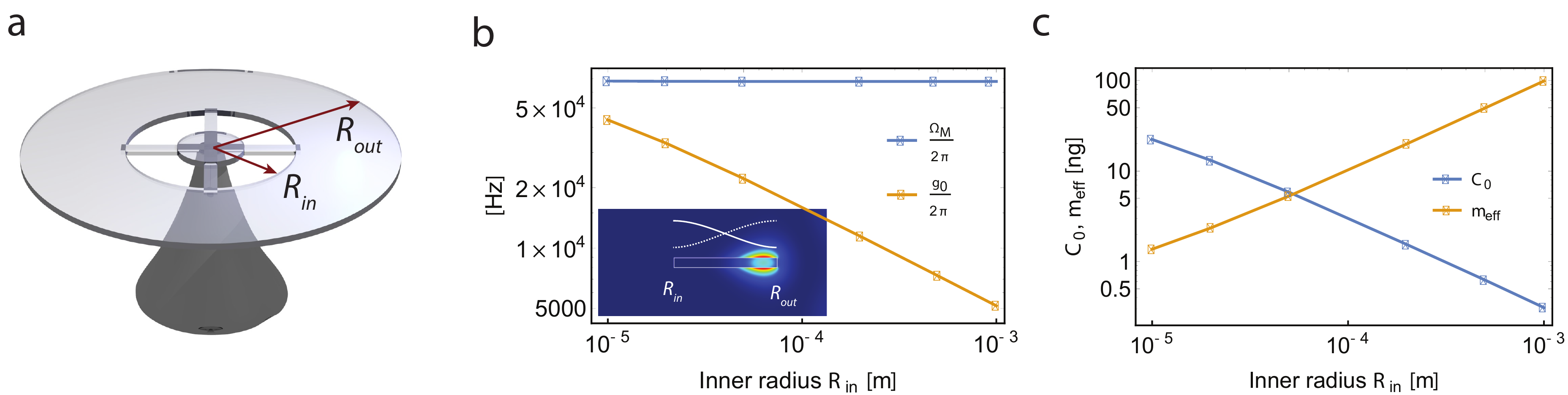}
\caption{\textbf{Annular disk case}. (a) Schematic of an annular disk resonator, of inner radius $R_{in}$ and outer radius $R_{out}$. (b) Calculated single photon optomechanical coupling strength $g_0/2\pi$ and mechanical frequency $\Omega_M/2\pi$ for the ($m$=0; $n$=1) third sound mode with free-free boundary conditions (see inset) as a function of $R_{in}$, for a $d=30$ nm thick superfluid film. The value of $R_{out}-R_{in}$ is kept constant at 4 microns. We obtain $G/2\pi=4.6$ GHz/nm (accounting for the mode overlap between third sound profile and WGM) from FEM simulations (see inset). (c) Predicted single photon optomechanical cooperativity $C_0$ and effective mass $m_{\mathrm{eff}}$ as a function of inner radius, for the same parameters as (b).}
\label{Figure5fig}
\end{figure}

Here we consider the case of a third sound wave confined on the surface of an annular disk resonator, as illustrated in Fig. \ref{Figure5fig}(a). In this case, the surface deformation profile $\eta_{m,n}\left(r,\theta \right)$ is given by:
\cite{gottlieb_harmonic_1979}
\begin{equation}
\eta_{m,n}\left(r,\theta \right)= \left(A_{m,n}\,J_m\left(k_{m,n}\, r\right)+ B_{m,n}\,Y_m\left(k_{m,n}\, r\right) \right) \cos\left(m \theta\right),
\label{EqBesselannulus}
\end{equation}
where $m$ and $n$ are respectively the azimuthal and radial mode numbers, $A_{m,n}$ and $B_{m,n}$ mode amplitude coefficients and $J_m$ and $Y_m$ respectively the Bessel functions of the first and second kind of order $m$.
For the free-free boundary condition in $r=R_{in}$ and $r=R_{out}$, the wavenumber $k_{m,n}$ is defined as the $n^{th}$ root of the equation:
\begin{equation}
J_m'\left( k\, R_{out} \right)\,Y_m'\left( k \, R_{in} \right)-Y_m'\left( k\, R_{out} \right)\,J_m'\left( k \,R_{in} \right)=0,
\end{equation}
and the coefficients A and B are found by imposing the free boundary conditions in $R_{in}$ and $R_{out}$ (i.e. $\partial_r \eta(R_{in})=\partial_r \eta(R_{out})=0$). Following the same approach outlined in the main text, we numerically calculate the values of $\Omega_M/2\pi$ and $g_0/2\pi$ as a function of resonator dimensions for the annular disk case. This is shown in Fig.  \ref{Figure5fig}(b), where $R_{in}$ is swept from $10^{-5}$ m to 1 mm, while the annulus width ($R_{out} - R_{in}$) is kept constant at 4 microns. As the third sound frequency only depends on the annular width, $\Omega_M$ remains constant over the entire range (blue line). 
For a fixed annular width, increasing $R_{in}$ results in the effective mass increasing nearly linearly (simply as the resonator surface area), resulting in a much less dramatic decrease in $g_0$ with resonator radius when compared to the disk case (see Fig. \ref{Figure4fig}), with for example $g_0>2\pi\times 5$ kHz on millimeter-sized annular resonators. Fig.  \ref{Figure5fig}(c) plots the dependence of $C_0$ and $m_{\mathrm{eff}}$ over the same parameter range. The predicted value of $C_0$ assumes a constant $Q_M=4000$ and $Q_{\mathrm{opt}}=10^7$, as in the disk case (see section \ref{sectionInfluenceoffilmthickness}), and reaches 22 for the smallest resonators.

\section{Appendix B: Microsphere resonator}
\label{sectionmicrosphereresonator}

\begin{figure}
\centering
\includegraphics[width=\textwidth]{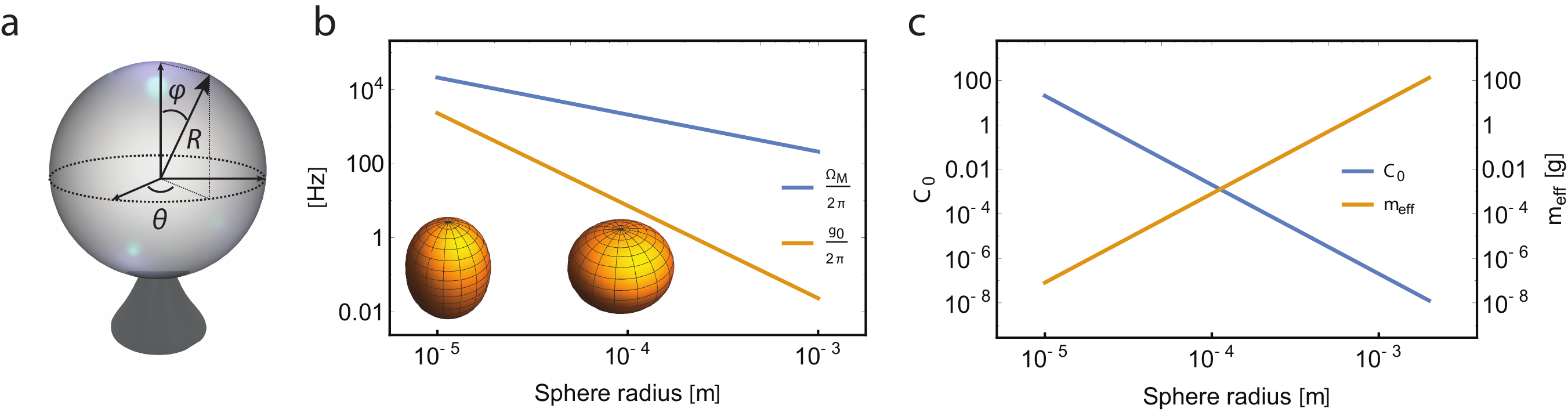}
\caption{\textbf{Microsphere case}. (a) Schematic of a microsphere on a pedestal, with spherical coordinates $R$, $\theta$ and $\varphi$. (b) Single photon optomechanical coupling strength $g_0/2\pi$ and mechanical frequency $\Omega_M/2\pi$ for the ($l$=2; $m$=0) third sound mode (see inset) as a function of sphere radius, for a $d=30$ nm thick superfluid film. Solid lines: analytical formula. (c) Predicted single photon optomechanical cooperativity $C_0$ and effective mass $m_{\mathrm{eff}}$ as a function of sphere radius, for the same third sound mode and film thickness as (b).}
\label{Figure6fig}
\end{figure}

Here we briefly address the optomechanical coupling between third sound and light respectively confined on the surface of and inside a microsphere resonator \cite{vernooy_high-q_1998, matsko_optical_2006}, see Fig. \ref{Figure6fig}(a). In analogy to the disk case, we model the third sound mode profile $\eta_{l,m}$ on a sphere as:
\begin{equation}
\eta_{l,m}\left(\theta,\varphi\right)=A_{l,m}\, Y_{l}^{m} \left(\theta,\varphi\right)
\label{Eqetasphere}
\end{equation}
with $A_{l,m}$ the mode amplitude and $Y_{l}^{m}$ the Laplace spherical harmonic of degree $l$ and order $m$, solution to Laplace's equation on a sphere.  Following the same treatment outlined in section \ref{subsectioneffectivemass}, we derive the effective mass of a point situated on the sphere's equator ($\varphi=\pi/2$):
\begin{equation}
m_{\mathrm{eff,\,sphere}}=\frac{2 \rho\,\int_{\varphi=0}^{\pi} \int_{\theta=0}^{2\pi}\left(\int_d^{d+\eta \left(  \theta,\varphi\right)} U\left(z\right)\rmd z\right)  R^2\, \rmd \theta\, \sin\left( \varphi \right)\rmd \varphi}{\eta^2\left(\theta, \,\varphi=\pi/2 \right)\,\Omega_M^2}\,.
\label{Eqmeffsphere}
\end{equation}
Which, substituting $\Omega_M=c_{3}\sqrt{l(l+1)}/R$, for a  mode rotationally invariant along $\theta$ simplifies to:
\begin{equation}
m_{\mathrm{eff,\,sphere}}=\left(\frac{\rho}{\rho_s}\right) \, \frac{R^4}{d} \, \frac{ 2\pi\rho}{l\left(l+1\right)}\, \frac{\int_{\varphi=0}^{\pi}\eta^2\left( \varphi \right)\sin\left( \varphi \right) \rmd \varphi}{\eta^2\left(\varphi=\pi/2 \right)}\,.
\end{equation}
Here we recognize the same $R^4/d$ effective mass scaling as in the disk case (see Eq. \ref{Eqmeffsuperfluid}). Next, as discussed in section \ref{section1optics}, we set $G_{\mathrm{sphere}}=\frac{\omega_0}{R}\left(\frac{1-\varepsilon_\mathrm{sf}}{1-\varepsilon_\mathrm{SiO_2}}\right)$
and use this approximation to calculate the optomechanical coupling rate $g_0$. This is shown in Fig. \ref{Figure6fig}(b) for the ($l$=2; $m$=0) third sound mode. This mode corresponds to the superfluid sloshing back and forth between the equator and the poles, as illustrated in the inset,  and  naturally exhibits good optomechanical coupling to a WGM localized on the sphere's equator. As in the case of a disk resonator, $\Omega_M$ scales as 1/$R$ and $x_{\mathrm{ZPF}}$ scales as $R^{-3/2}$; however the dependence of $g_0$ on $R$ is even steeper than in the disk case, since for a sphere $G$ is also inversely proportional to $R$. For a sphere of radius $R=20$ microns and a 30 nm thick film,  $g_0= 2\pi \times 410$ Hz, approximately 28 times less than for the ($m$=0; $n$=1) mode of a disk of identical radius (see Fig. \ref{Figure4fig}(b)). This smaller value has two distinct origins: a $2.2$ times smaller $x_{\mathrm{ZPF}}$ for the third sound mode on the sphere, because of its larger effective mass, and a $12.5$ times smaller G compared to the disk case because of the weaker interaction between the superfluid film and the WGM in the sphere, as discussed in section \ref{section1optics}. Finally Fig. \ref{Figure6fig}(c) plots the dependence of $C_0$ and $m_{\mathrm{eff}}$ on sphere radius, for a 30 nm thick superfluid film. The predicted value of $C_0$ assumes a constant $Q_M=4000$ and $Q_{\mathrm{opt}}=10^9$ \cite{vernooy_high-q_1998}. This plot underscores the strong dependence of $m_{\mathrm{eff}}$ with sphere radius: as R goes from 10 microns to 2 mm, $m_{\mathrm{eff}}$ spans over 9 orders of magnitude. For $R=2$ mm, $m_{\mathrm{eff}}$  reaches 130 g, i.e. over $5\times 10^8$ times the actual mass of the superfluid film.




\bibliographystyle{unsrt} 
\bibliography{referencessuperfluid}

\end{document}